# ARE CHROMOSPHERIC NANOFLARES A PRIMARY SOURCE OF CORONAL PLASMA?

J. A. Klimchuk

NASA Goddard Space Flight Center, Greenbelt, MD 20771, USA;
James.A.Klimchuk@nasa.gov

S. J. Bradshaw

Department of Physics and Astronomy, Rice University, Houston, TX 77005, USA;
stephen.bradshaw@rice.edu

## ABSTRACT

It has been suggested that the hot plasma of the solar corona comes primarily from impulsive heating events, or nanoflares, that occur in the lower atmosphere, either in the upper part of the ordinary chromosphere or at the tips of type II spicules. We test this idea with a series of hydrodynamic simulations. We find that synthetic Fe XII (195) and Fe XIV (274) line profiles generated from the simulations disagree dramatically with actual observations. The integrated line intensities are much too faint; the blue shifts are much too fast; the blue-red asymmetries are much too large; and the emission is confined to low altitudes. We conclude that chromospheric nanoflares are not a primary source of hot coronal plasma. Such events may play an important role in producing the chromosphere and powering its intense radiation, but they do not, in general, raise the temperature of the plasma to coronal values. Those cases where coronal temperatures are reached must be relatively uncommon. The observed profiles of Fe XII and Fe XIV come primarily from plasma that is heated in the corona itself, either by coronal nanoflares or a quasi-steady coronal heating process. Chromospheric nanoflares might play a role in generating waves that provide this coronal heating.

## 1. INTRODUCTION

All hot (~$10^6$ K) coronal plasma comes originally from the cold (~$10^4$ K) chromosphere, but the scenario by which this occurs is still being debated. In the standard picture, magnetic free energy or wave energy is dissipated above the chromosphere and heats the plasma locally to high temperatures. The coronal density may be very small at this time. Most of the energy is transported downward by thermal conduction and deposited in the transition region and upper chromosphere. Particle beams can reach even deeper layers. This heats the plasma, raising its temperature and pressure, and causing it to flow upward by the well-known process of chromospheric evaporation. The corona becomes filled with hot evaporated plasma. If the coronal energy input persists, a static equilibrium is eventually achieved in which radiative and conductive cooling exactly balance the input. In the transition region, the energy balance is between radiation and the downward thermal conduction flux. When the coronal heating eventually ceases, the plasma cools and drains back down to the chromosphere. (See the discussions in Klimchuk et al. 2008 and Reale 2010 for more details.)

Several lines of evidence suggest that coronal heating may be very impulsive, with the energy release only lasting for a short time (see Klimchuk 2006, 2009; Reale 2010, and references cited therein). In that case the evaporated plasma cools and drains immediately, only to be evaporated again in a subsequent heating event. Such impulsive, small spatial scale heating events are often called nanoflares. We use the term generically. No specific physical mechanism is implied. Possible candidates include magnetic reconnection, secondary instability, kink instability, turbulence, and resonant wave absorption (Klimchuk 2006; Parnell & de Moortel 2012). The frequency with which nanoflares repeat on a given magnetic strand has a strong influence on the thermal distribution of the plasma and has become a topic of great interest (e.g., Bradshaw et al. 2012; Warren et al. 2012; Reep et al. 2013; Cargill 2014).

A completely different scenario for the origin of hot plasma proposes that nanoflares occur in the upper chromosphere rather than in the corona (Hansteen et al. 2010). The plasma is heated directly rather than by a thermal conduction flux. Because of the sudden increase in pressure, the hot plasma expands rapidly upward into the corona. Depending on the impulsiveness of the event, this expansion can be very explosive. There is some downward expansion as well, but this is largely resisted by the strong equilibrium pressure gradients of the deeper layers.

One interpretation of type II spicules involves what can be thought of as chromospheric nanoflares. Like the more common type I (classical) spicules, type II's are upward jets of $10^4$ K plasma. Whereas type I's maintain this temperature as they rise and fall, type II's are heated as they are ejected. Most of the material reaches a temperature of only $10^5$ K before falling back to the surface, but a fraction (~10%) at the tip is observed in coronal emission lines and continues to rise upward (De Pontieu et al. 2011). One possibility is that this hot emission comes *not* from ejected material, but rather from pre-existing coronal plasma that is shock heated as the jet plows into it (Klimchuk 2012; Petralia et al. 2014). Type I spicules are not fast enough to produce shocks, but many type IIs are. The other possibility is that the hot material seen at the tip is indeed part of the spicule itself. Its heating must be impulsive, since the rising jet disappears in cool emissions on a time scale of only 5-20 s (De Pontieu et al. 2007). In this scenario, the spicule introduces new hot material into the corona that was not there before. This led to the reasonable suggestion that type II spicules might provide a substantial fraction of the total mass that exists in the corona (De Pontieu et al. 2009, 2011; Moore et al. 2011).

Klimchuk (2012; henceforth K12) recently investigated this suggestion. He identified three observational tests of the hypothesis that all of the hot plasma in the corona comes from type II spicules:  the ratio of blue wing to line core intensities in coronal emission lines, the ratio of lower transition region ($\leq 10^5$ K) to coronal emission measures, and the ratio of densities measured in the blue wing and line core of density-sensitive coronal lines. It was found that the hypothesis fails all three tests by a wide margin (K12; Patsourakos et al. 2014; Tripathi & Klimchuk 2013). Observed ratios differ from the predicted ratios by roughly two orders of magnitude. It was concluded that only a small fraction of the plasma in the corona can come from the heated tips of spicules.

The work reported here is in many ways a follow-up to K12. It validates with numerical simulations the analytical results of that study. Equally importantly, it treats chromospheric nanoflares as a general phenomenon, not necessarily associated with spicules. The simulations impose an impulsive energy release in the upper chromosphere of an initially static atmosphere. No ejection of cool material is assumed, nor is one produced. The results are nonetheless relevant to the tips of spicules, because the dynamics of the heated plasma are dominated by its explosive expansion. The observed ejection velocity of type II spicules, ~100 km s$^{-1}$, is slow compared to the expansion velocities in our simulations.

As in K12, we compute spectral line profiles assuming that essentially all coronal plasma comes from chromospheric nanoflares. We use the well observed lines of Fe XII (195.119 A) and Fe XIV (274.204 A), which are formed at 1.5 and 2.0 MK, respectively, under conditions of ionization equilibrium. The plasma is out of equilibrium, however, and we therefore employ a full nonequilibrium ionization (NEI) treatment. We measure the intensities, Doppler shifts, and blue-red (BR) asymmetries of the lines, and compare them to actual observations. From the comparison, we are able to draw a strong conclusion about the role of chromospheric nanoflares in explaining the corona. Details and discussion are provided in the following sections.

## 2. NUMERICAL MODEL

MHD processes are no doubt responsible for chromospheric nanoflares, but once the heating has occurred, the subsequent evolution of the plasma is controlled primarily by magnetic field-aligned processes. A one-dimension hydrodynamic approach is therefore appropriate. Some lateral expansion of the heated flux tube may occur if the plasma β approaches unity (not expected for the pressures in our simulations, especially with active region field strengths), but most of the motion will be in the direction of least resistance, i.e., upward into the low-pressure ambient corona. We model the evolution with the state-of-the-art HYDRAD code described in detail in Bradshaw & Cargill (2013). Two of its features are especially important: adaptive mesh refinement (AMR) and nonequilibrium ionization. AMR is necessary because of the steep gradients that are present. The code resolves these gradients by refining down to a cell size of 3.9 km in the experiments presented here. NEI effects can be strong because of the short timescales involved in the rapid heating of the plasma and its subsequent rapid cooling due to expansion.

We imagine that randomly occurring chromospheric nanoflares populate a magnetic arcade with plasma. A line-of-sight through the optically thin plasma will intersect many different magnetic strands and therefore sample plasma from many different events. In addition, spectroscopic observations typically have modest spatial resolution (~ 2 arcsec) and rather long integration times (several 10s of seconds), so a typical observation will represent a spatial and temporal average. We approximate this by spatially and temporally averaging the simulation of a single representative strand, as indicated schematically in Figure 1. It is as though the strand is broken up into segments that are vertically aligned along the line-of-sight. We average over both halves of the strand (not indicated in the figure), since nanoflares are assumed to be equally likely at the two footpoints. For two of our simulations (Runs 1 and 3), we choose a semi-circular strand of length 157 Mm. The apex height is 50 Mm, which equals the characteristic coronal height used in K12. For a third simulation (Run 2), we use a shorter strand of length 30 Mm.

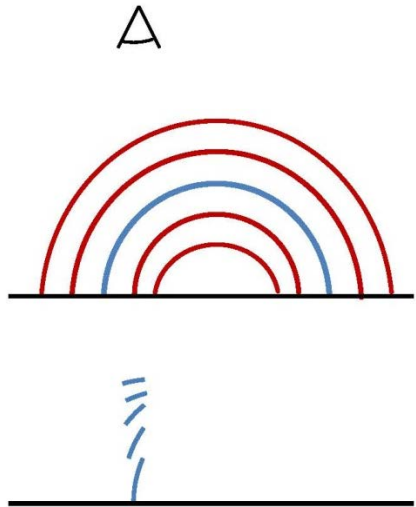

**Figure 1.** Schematic drawing showing how the spatially averaged emission from a single magnetic strand is representative of a line-of-sight passing through an arcade of many strands.

The bottom 5 Mm of each leg contains our model chromosphere. It is designed to contain many scale heights of plasma so that the evolution is not affected by the rigid wall boundary conditions at the base. The chromosphere is maintained at a nearly constant temperature of $10^4$ K (before the nanoflare) using the standard technique of modifying the optically thin radiative loss function so that it decreases precipitously to zero below the desired chromospheric temperature (e.g., Klimchuk et al. 1987). Although our chromosphere is idealized, it has all of the physical properties necessary to address the objectives of this study.

Since we wish to test the hypothesis that the corona is produced by chromospheric nanoflares, we assume that there exists only a low level of "background" heating. We adopt a steady uniform heating rate of $5.77 \times 10^{-7}$ erg cm$^{-3}$ s$^{-1}$ in the long strand and $5.09 \times 10^{-6}$ erg cm$^{-3}$ s$^{-1}$ in the short strand, which produce static equilibrium apex temperatures of 0.46 MK and 0.24 MK, respectively.

Starting with this initial equilibrium, we impose a nanoflare in the upper chromosphere of the "left" leg. It has a Gaussian spatial profile that is centered at spatial coordinate $s = 4.5$ Mm and has a full width of 0.5 Mm. The nanoflare lasts 10 s in Runs 1 and 2 and has the temporal profile of a top hat, ramping up to a peak heating rate of 1.5 erg cm$^{-3}$ s$^{-1}$ in the first second and back down to zero in the last second. For Run 3, we impose a more gradual 100 s nanoflare with a triangular temporal profile peaking at 0.6 erg cm$^{-3}$ s$^{-1}$. The magnitudes of the nanoflares were chosen to raise the temperature of the chromospheric plasma to approximately 2 MK, the characteristic temperature of the corona that we are attempting to explain. The parameters of the three simulations are given in Table 1.

Table 1.  Simulation Parameters

| Run | Strand Length (Mm) | Nanoflare Duration (s) | Nanoflare Energy (erg cm$^{-2}$) |
|---|---|---|---|
| 1 | 157 | 10 | $1.20 \times 10^9$ |
| 2 | 30 | 10 | $1.20 \times 10^9$ |
| 3 | 157 | 100 | $2.67 \times 10^9$ |

We run the simulation for 5000 s, which is enough time for the strand to return approximately to its initial pre-nanoflare equilibrium. At each second in the simulation, we compute the Fe XII (195) and Fe XIV (274) emissivities in each numerical cell. Knowing the emissivity, temperature, and line-of-sight velocity, we then construct a Gaussian line profile for the cell. We combine the profiles for the entire strand and for the full duration of the simulation to obtain a spatially and temporally averaged profile for the run. This represents the observation of a single pixel on the Sun. Two profiles are computed in each case:  one that assumes ionization equilibrium and one that uses the full nonequilibrium ionization treatment. All profiles and measured profile parameters presented here use NEI, but the basic conclusions are the same with equilibrium ionization.

3.  RESULTS

Figure 2 shows electron temperature (solid), electron number density (dotted), and velocity (dashed) as a function of position along the strand at times of 0, 10, 30, and 60 s in Run 1. Note that density is plotted logarithmically, with 0 corresponding to $\log(n) = 6.0$ in the initial equilibrium and log($n$) = 7.5 at the other times. Note also that the spatial range is different in the four plots.

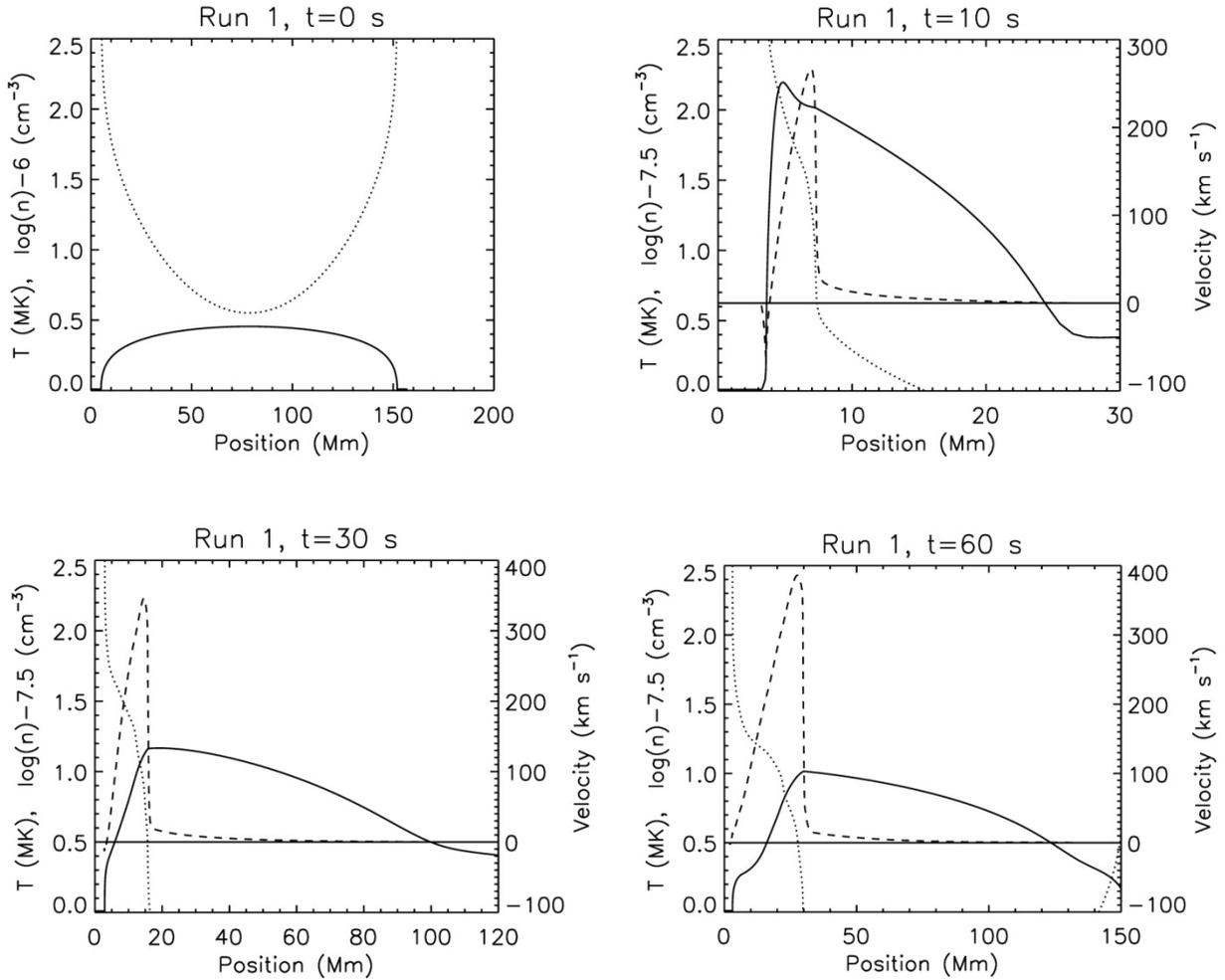

**Figure 2.** Temperature (solid), electron density (dotted), and velocity (dashed) as a function of position along the strand at $t$ = 0, 10, 30, and 60 s in Run 1. Density is plotted logarithmically, with a different offset at $t$ = 0 s.

Several interesting effects are apparent. First, thermal conduction efficiently raises the temperature of the tenuous corona along the entire strand. Only a small bump at $s$ = 4.5 Mm in the 10 s temperature plot reveals the Gaussian heating profile of the nanoflare. Second, the high pressure plug produced by the nanoflare expands explosively upward, with velocities approaching 400 km s$^{-1}$. These flows are supersonic, so an isothermal shock develops at the leading edge. (The plug extends to where the velocity and density profiles have a sharp drop off.) Third, both the density and temperature of the plug quickly decrease as the plasma expands. By $t$ = 60 s, the density has fallen by about a factor of 10 and the temperature has fallen by about a factor of 3 compared to the end of the nanoflare at $t = 10$ s. The large density decrease is of course expected from the rapid increase in volume. The large temperature decrease is also expected, since, as discussed in K12, the work done by adiabatic expansion produces strong cooling, even in the absence of thermal conduction and radiation losses. The importance of adiabatic cooling is often under appreciated.

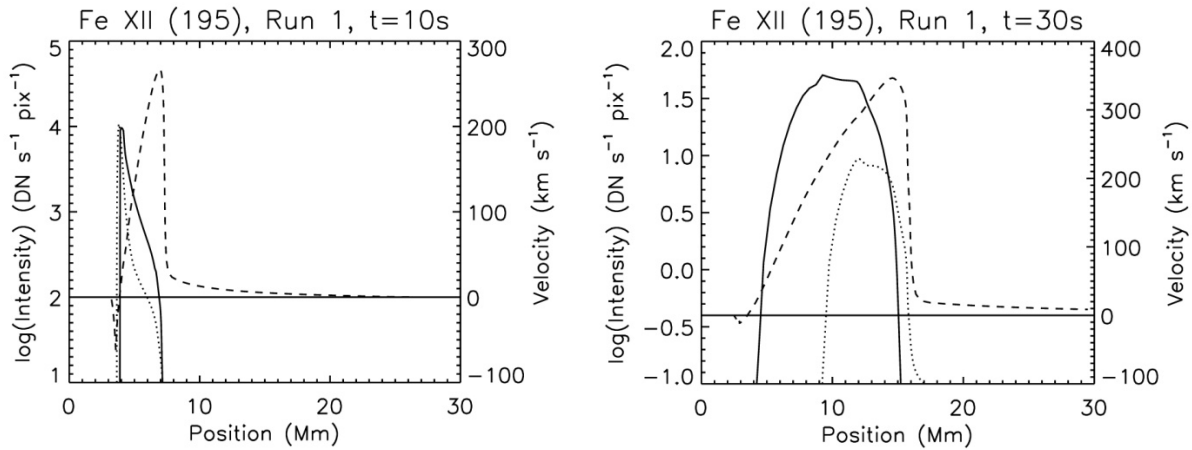

**Figure 3.** Intensity of Fe XII (195) assuming nonequilibrium (solid) and equilibrium (dotted) ionization and velocity (dashed) as a function of position along the strand at $t = 10$ and 30 s in Run 1. Intensity is plotted logarithmically.

Figure 3 shows the Fe XII (195) emissivity as a function of position along the strand at $t = 10$ and 30 s. It is presented as an intensity corresponding to the count rate in data numbers (DN) that would be measured by the Extreme ultraviolet Imaging Spectrometer (EIS) on Hinode if there were a uniformly filled plasma column having a length (line-of-sight depth) of 50 Mm and cross section equal to an EIS pixel. The solid curve is for NEI, and the dotted curve is for equilibrium ionization. Velocity is indicated by the dashed curve, as in Figure 2. Taken together, the intensity and velocity curves are useful for understanding the contributions to the different wavelength positions in the spectral line profile.

The right side of the intensity curve ($s > 4$ Mm at $t = 10$ s) is determined primarily by the spatial dependence of emission measure, since temperature is roughly constant in the heated plug and corona above. Thus, the intensity curve mimics the density curve in Figure 2. The left side of the intensity curve is instead determined by temperature, which plummets below the plug. Thermal conduction can easily raise the temperature of the low density corona above the plug, but it is far less effective in the high density chromosphere below.

The NEI emission is stronger than the equilibrium emission at both times shown in Figure 3. In the first case ($t = 10$ s), it is because ionization cannot keep up with the rapid temperature increase during the heating phase. Fe XII is more abundant than it normally would be at these temperatures, which exceed the 1.5 MK temperature of maximum abundance under equilibrium conditions. In the second case ($t = 30$ s), recombination cannot keep pace with the rapid cooling from expansion, and Fe XII is abundant even though the temperatures have dropped below 1.5 MK.

The intensity decreases rapidly as the plasma expands, due partly to the rapid decrease in density and partly to the rapid decrease in temperature. Count rates fall by two orders of magnitude between $t = 10$ and 30 s and by another order of magnitude between $t = 30$ and 60 s. This has two important consequences. First, the emission is confined to low altitudes in the corona (< 10 Mm above the chromosphere), contrary to observations. Second, the temporally-averaged emission is very faint, also contrary to observations.

Figure 4 shows the composite Fe XII (195) and Fe XIV (274) line profiles for the entire simulation (using NEI). These are the profiles obtained by averaging in both space and time. Although they come from a single strand, they approximate an observation of many unresolved strands in different stages of evolution, as discussion in Section 2. We have again normalized assuming a filled plasma column having a length (line-of-sight depth) of 50 Mm and cross section equal to an EIS pixel, but the plasma is not uniform, as assumed for the intensity versus position plots at specific times in Figure 3. Instead, the column has a mixture of temperatures and densities corresponding to the spatial and temporal average of the full simulation. The total intensity integrated over the Fe XII line profile is 0.15 DN s$^{-1}$ pix$^{-1}$. In comparison, EIS observes typical intensities of 250 DN s$^{-1}$ pix$^{-1}$ in active regions (Tripathi et al. 2011; Warren et al. 2012[1]) and 25 DN s$^{-1}$ pix$^{-1}$ in quiet Sun (Brooks et al. 2009;

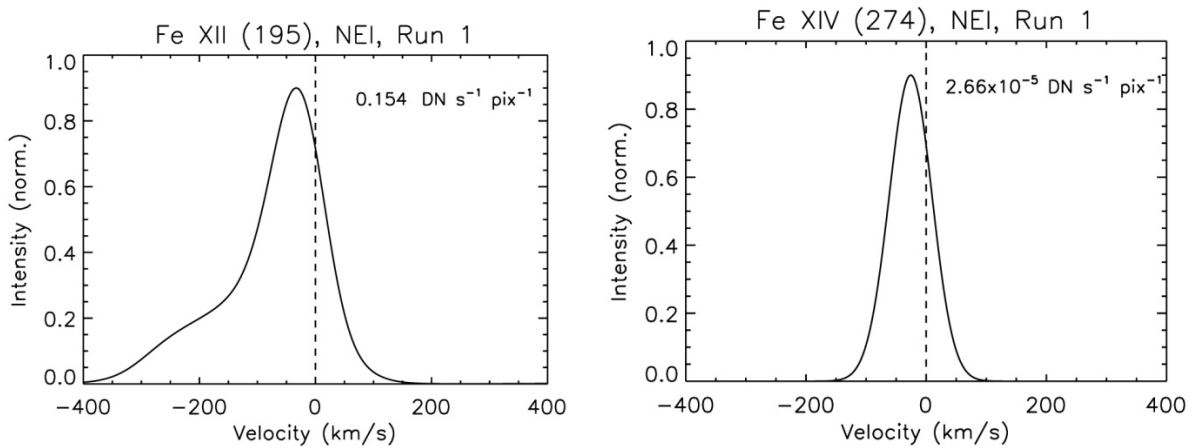

**Figure 4.** Spatially and temporally averaged profiles of Fe XII (195) and Fe XIV (274) for Run 1.

---

[1] Fe XIV (274.203) intensities were estimated from Fe XIV (270.519) intensities using a factor of 2 conversion that is appropriate for a density of $10^9$ cm$^{-3}$ (H. Warren, private communication).

Mariska 2013). The total intensity of Fe XIV is $2.7 \times 10^{-5}$ DN s$^{-1}$ pix$^{-1}$, which is essentially invisible, compared to typical observed values of 50 and 1 DN s$^{-1}$ pix$^{-1}$ in active regions and quiet Sun, respectively. Thus, the predicted intensities are several orders of magnitude too small.

The shape of the Fe XII profile presents another major discrepancy. There is a strong enhancement of the blue wing that is not observed. We quantify this with a parameter called the blue-red (or red-blue) asymmetry. It is a measure of the excess blue wing emission normalized by the line core emission. We compute BR asymmetries by integrating over the blue (-250 to -50 km s$^{-1}$) and red (+50 to +250 km s$^{-1}$) wings of the profile and over the line core (-30 to +30 km s$^{-1}$). The asymmetry is the difference between the blue and red intensities divided by the core intensity. It is defined to be positive when the blue wing is brighter. The BR asymmetry of the Fe XII profile in Figure 4 is 0.65 when we use the wavelength of peak intensity for the "rest" wavelength, $\lambda_0$, as is usually done in observational studies. The asymmetry is 1.6 if we instead use the actual rest wavelength. The Fe XIV profile has an asymmetry of 0.51 using the actual rest wavelength and almost no asymmetry using $\lambda_0 = \lambda(I_{max})$.

Observed BR asymmetries are generally < 0.05 in active regions, quiet Sun, and coronal holes (Hara et al. 2008; De Pontieu et al. 2009, 2011; McIntosh & De Pontieu 2009; Tian et al. 2011; Doschek 2012; Patsourakos et al. 2014; Tripathi & Klimchuk 2013). "Outflow regions" at the periphery of active regions are an exception. The asymmetry there can reach and sometimes exceed 0.2. These outflow regions are faint areas that may be the source of slow solar wind (van Driel-Gesztelyi et al. 2012).

K12 estimated BR asymmetries of generally > 1 using the actual rest wavelength for $\lambda_0$. However, he assumed that the line core emission is produced during a slow cooling and draining phase after the heated plasma has expanded to fill the strand. In order for this plasma to be visible in a coronal line, it must still be hot at the end of the expansion phase. K12 emphasized that this implies continued strong heating in the corona after the nanoflare has ended, to counteract the cooling effect of the expansion. Also, K12 considered nanoflares that occur at the tips of type II spicules, when the plasma is already moving at ~ 100 km s$^{-1}$. In that case, the emission is strongly blue shifted even early in the nanoflare energy release. In our case, the heated plasma is initially at rest, and substantial emission is produced at low velocity. This can be seen by comparing the intensity and velocity curves at $t = 10$ s in Figure 3. It is this emission that produces the line core, not emission from a slow draining phase after the strand has been filled. Had we started with an upflow rather than a static equilibrium, our profiles would be shifted to the blue by an amount equal to the initial velocity.

We measure the Doppler shifts of our synthetic line profiles using the center of mass (first moment) technique (Dere et al. 1984). Fe XII is blue-shifted by 74 km s$^{-1}$ and Fe XIV is blue shifted by 26 km s$^{-1}$ relative to the actual rest wavelengths. Observed Doppler shifts of coronal lines are generally slower than 5 km s$^{-1}$ in active regions (Doschek 2012; Tripathi et al. 2012) and quiet Sun (Chae et al. 1998; Peter & Judge 1999) and can be blue or red, while coronal holes exhibit blue shifts of approximately 10 km s$^{-1}$ (Rottman et al. 1982). Since few observations have an absolute wavelength calibration, these shifts are often measured relative to a raster average or some other estimate of the true zero.

The integrated line intensity, blue shift, and BR asymmetries predicted from Run 1 are given in the first row of Tables 2 and 3 for Fe XII and Fe XIV, respectively. Typical observed values are given in the bottom row. The discrepancies are large, except for the BR asymmetry of Fe XIV.

Run 2 uses the same 10 s nanoflare as Run 1, but in a much shorter strand of length 30 Mm (20 Mm between the tops of the chromospheres at the two ends). Run 3 uses the long strand, but imposes a longer nanoflare lasting 100 s. Figures 5 and 6 show the composite Fe XII and Fe XIV profiles for these two simulations. The profile parameters are given in the second and third rows of Tables 2 and 3. Although the line intensities are brighter than in Run 1, they are still approximately two orders of magnitude fainter than observed, especially in active regions. The blue shifts are again much faster than observed, and the line asymmetries are much larger than observed for both choices of $\lambda_0$. The Fe XIV profile for Run 2 has the interesting property of being so strongly blue shifted (109 km s$^{-1}$) that the asymmetry is actually negative when $I_{max}$ is used for $\lambda_0$. Negative asymmetry generally suggests an excess of red emission, but in this case nearly all of the emission is blue shifted. The Fe XIV profile for Run 3 has a double peak, which is rarely observed.

## 3. DISCUSSION

All three of our chromospheric nanoflare simulations predict Fe XII (195) and Fe XIV (274) line profiles that are grossly inconsistent with observations. The line intensities are too faint, the blue shifts are too fast, and the BR asymmetries are too large, all by one or two orders of magnitude. In addition, the emission is confined to low altitudes in the corona. Clearly, the observed coronal emission lines cannot be produced primarily by this process. Chromospheric nanoflares of the types we have considered can play only a minor role in supplying the corona with hot plasma. Our simulations apply specifically to nanoflares that occur in an equilibrium chromosphere, in the absence of a type II spicule, but the blue shifts and BR asymmetries would be even more discrepant for impulsive heating at the tip of a moving spicule. Our results therefore validate the conclusions in K12.

It is possible to increase the predicted line intensities, though the other discrepancies would remain. Our composite profiles average over the full 5000 s of the simulation, which is equivalent to assuming that nanoflares repeat in each strand with a delay of this length. Shorter delays would give rise to brighter emission. For example, reducing the delay to 2000 s would increase the intensities by a factor of 2.5. This is still far less than what is needed to explain the observations. Shorter delays would increase the intensities further, but they would not provide enough time for the plasma to return to an approximate equilibrium, which would be inconsistent with the initial conditions of our simulations. Furthermore, rapidly repeating chromospheric nanoflares present other observational challenges, as discussed in K12. We have nonetheless begun to simulate this scenario and will report on the results in due course.

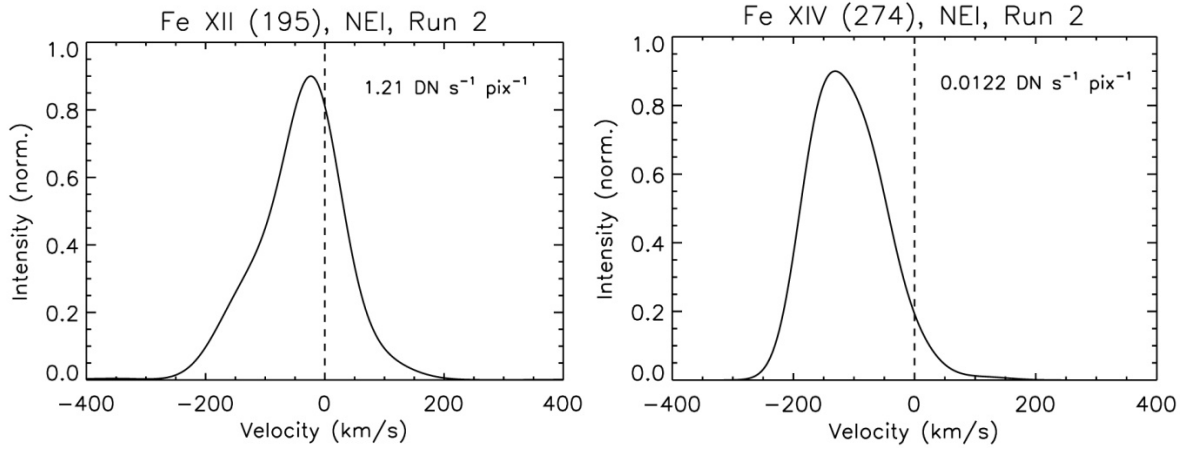

**Figure 5.** Spatially and temporally averaged profiles of Fe XII (195) and Fe XIV (274) for Run 2.

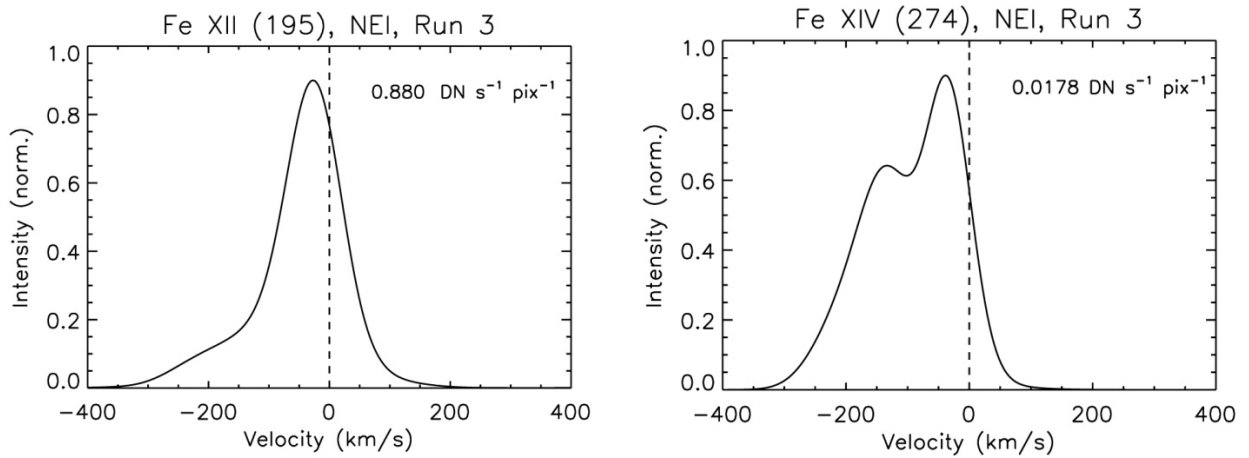

**Figure 6.** Spatially and temporally averaged profiles of Fe XII (195) and Fe XIV (274) for Run 3.

Table 2.  Fe XII (195) Parameters

| Run | $I_{line}$ (DN s$^{-1}$ pix$^{-1}$) | Blue Shift (km s$^{-1}$) | BR Asymmetry $\lambda_0 = \lambda(I_{max})$ | BR Asymmetry True $\lambda_0$ |
|---|---|---|---|---|
| 1 | 0.15 | 74 | 0.65 | 1.6 |
| 2 | 1.2 | 41 | 0.38 | 0.99 |
| 3 | 0.88 | 49 | 0.32 | 1.0 |
| Observed | 25-250 | < 10 | < 0.05 | |

Table 3.  Fe XIV (274) Parameters

| Run | $I_{line}$ (DN s$^{-1}$ pix$^{-1}$) | Blue Shift (km s$^{-1}$) | BR Asymmetry $\lambda_0 = \lambda\,(I_{max})$ | BR Asymmetry True $\lambda_0$ |
|---|---|---|---|---|
| 1 | 2.7x10$^{-5}$ | 26 | 7.0x10$^{-3}$ | 0.51 |
| 2 | 0.012 | 109 | -0.52 | 9.9 |
| 3 | 0.018 | 94 | 1.4 | 3.1 |
| Observed | 1-50 | < 10 | < 0.05 | |

The predicted intensities would also be brighter if the upper chromosphere in the initial equilibrium were more dense. This is possible, but only if the density is higher everywhere in the strand, including the corona, which requires a stronger background coronal heating rate. We can estimate the relationship between intensity and background heating rate in the following way. For simplicity, let us assume that the upper chromosphere is heated instantaneously by the nanoflare. Its brightness is then proportional to the square of its initial density, $I \propto n_{ch}^2$. For a given pre-event chromospheric temperature, the initial density is proportional to the initial pressure, $n_{ch} \propto P_{ch}$. But in equilibrium, the pressure at the top of the chromosphere is approximately equal to the coronal pressure, $P_{ch} \approx P$, and the coronal pressure is related to the coronal heating rate through the scaling law $P \propto Q^{11/14}$ (Porter & Klimchuk 1995). We have used the fact that the optically thin radiative loss function is approximately constant in the range $0.4 < T$ (MK) $< 1.5$ (Klimchuk et al. 2008). We therefore have that the intensity of the impulsively heated chromospheric plasma depends on the background coronal heating rate as $I \propto Q^{11/7}$. To increase the intensity by a factor of 100, bringing it more in line with observations, requires a factor of 19 increase of the heating rate. However, from another equilibrium scaling law, $T \propto Q^{2/7}$ (Porter & Klimchuk 1995), an increase of this magnitude in the heating produces a factor of 2.3 increase in the coronal temperature. This means starting with an initial corona that is hotter than 1 MK for Runs 1 and 3. This would be sufficient to explain the observed corona, and there would be no need for chromospheric nanoflares!

We emphasize that the blue shifts, BR asymmetries, and confinement of the emission to low altitudes would be largely unchanged if the nanoflares occurred in a higher density plasma. This is true whether the density of the upper chromophere is increased, as discussed above, or the nanoflare takes place in a deeper layer. In either case, a temperature increase from $10^4$ to $10^6$ K is accompanied by a locally enhanced pressure that is two orders of magnitude greater than the confining pressure above. An explosive upflow is unavoidable. Expansion cooling of this upflow is also unavoidable, independent of the density, so the line profiles would have similar properties and the large discrepancies with observations would remain.

We conclude that hot coronal plasma cannot come primarily from chromospheric nanoflares. This includes impulsive heating in the ordinary chromosphere and impulsive heating at the tips of type II spicules. We are *not* suggesting that chromospheric nanoflares do not occur or are not important. On the contrary, we feel they may play a vital role in heating the chromosphere and powering its intense radiation. However, these nanoflares do not, in general, raise the temperature to coronal values. Those cases where coronal temperatures are reached must be relatively uncommon. Otherwise, profiles of Fe XII and Fe XIV would be much different than observed. These profiles are dominated by emission from plasma that is heated in the corona itself, either by coronal nanoflares or a quasi-steady coronal heating process. Chromospheric nanoflares might play a role in generating waves that provide this coronal heating. Impulsive heating in the main bodies of type II spicules might also explain the bright emission and substantial red shifts at $T < 0.1$ MK that present a challenge to conventional models (K12).

This work was supported by the NASA Supporting Research and Technology Program. The authors benefited from participation in the International Space Science Institute team on Using Observables to Settle the Question of Steady vs. Impulsive Coronal Heating, led by one of us (SJB) and Helen Mason.